\documentclass[twocolumn]{emulateapj}
\usepackage{amssymb}

\begin{document}

\newcommand{\be}{\begin{equation}}
\newcommand{\ee}{\end{equation}}
\newcommand{\bc}{\begin{center}}
\newcommand{\ec}{\end{center}}

\title{The Challenge of Sub-Keplerian Rotation for Disk Winds}

\author{Frank H. Shu$^1$, Susana Lizano$^2$, Daniele Galli$^3$, 
Mike J. Cai$^4$, Subhanjoy Mohanty$^5$}

\affil{$^1$Department of Physics, University of California, San Diego, CA 
92093\\
$^2$CRyA, Universidad Nacional Aut\'onoma de M\'exico, 58089 Morelia, Mexico\\
$^3$INAF-Osservatorio Astrofisico di Arcetri, 
Largo E. Fermi 5, Firenze I-50125, Italy\\
$^4$Academia Sinica, Institute of Astronomy
and Astrophysics, Taiwan\\
$^5$Harvard-Smithsonian CfA, 60 Garden Street, Cambridge, MA 02138} 
\email{fshu@physics.ucsd.edu}

\begin{abstract}
\noindent Strong magnetization makes the disks surrounding young stellar objects
rotate at rates that are too sub-Keplerian to enable the thermal launching of disk
winds from their surfaces unless the rate of gas diffusion across field lines is
dynamically fast.  This under-appreciated implication of disk magnetization poses a considerable
challenge for disk-wind theory. 

\end{abstract}


\section{Introduction}

The pioneering study by Blandford \& Payne (1982, hereafter BP82; see
also Chan \& Henriksen 1980) opened the door to a physical
understanding of the highly collimated jets that emanate from
magnetized disks accreting onto central gravitating masses.  Pudritz \& Norman (1983) made the first
application to the disks in star formation, while Uchida \& Shibata (1985) performed
the first numerical simulations.  Collectively, the disk-wind community has contributed to much of what we know
about the magnetohydrodynamic (MHD) processes of jet launching, acceleration, and collimation.

Invoking strong magnetization to aid in the magnetohydrodynamics of the gas
above the disk has serious consequences for the
gas contained within the disk. Wardle \& K\"onigl (1993, hereafter WK93), 
Ferreira \& Pelletier (1995, FP95), Ferreira (1997, F97), Casse \& Ferreira (2000, CF00), and Salmeron, K\"onigl, \& Wardle (2007, SKW07) performed the primary studies
of some of the effects.  The result has been a
shrinkage of the viable parameter space to launch disk winds. In fact, unless the gas
diffuses across magnetic field lines at dynamical rates,
the sub-Keplerian rotation of the surfaces of realistic disks in young stellar objects (YSOs) may make magnetocentrifugally-driven disk winds a severe challenge if their surfaces are cold.  

In \S 2, we give the basic equations from the analysis of Shu et al. (2007, S07).  In \S 3, we use this analysis to demonstrate that fast disk winds that are more than lightly loaded will not arise if they are thermally launched.  In \S 4, we review how
WK93, FP95, F97, CF00, and SKW07 formally escape the conclusions of \S 3 by assuming that the diffusion of gas across magnetic field lines inside the disk occurs on dynamical time scales, and we discuss the astrophysical difficulties engendered by such a point of view.  In \S 5, we summarize how X-wind theory surmounts the same difficulties.

\section{Basic Equations}

Use a cylindrical coordinate system $(\varpi, \varphi, z)$ centered
on a star with mass $M_*$, and consider centrifugal balance in a thin
disk rotating at radius $\varpi$ with an angular velocity $\Omega(\varpi)$ that is a density-weighted average over $z$:
\be
- \Sigma \varpi\Omega^2 = {B_zB_\varpi^+ \over 2\pi} - {GM_*\Sigma\over \varpi^2},
\label{centrifugal} 
\ee
where $G$ is the universal gravitational constant, $\Sigma$ is the
surface density, and $B_z$ and $B_\varpi^+$ are,
respectively, the midplane and upper surface values of the vertical and
radial components of the magnetic field.
From Amp\`ere's law in cgs units, $cB_\varpi^+/2\pi$ equals the
vertically integrated current density $J_\varphi$, where $c$ is the
speed of light.  Equation (\ref{centrifugal}) then assumes
that the Lorentz force per unit area, $J_\varphi B_z/c$, associated
with magnetic tension, provides the only opposition, apart from
inertia, to the attraction of stellar gravity.  We have ignored the
self-gravity of the disk, which would additionally hinder the launching
of magnetocentrifugal winds.  When integrating the divergence of the Maxwerll stress tensor
over the disk thickness, we have also ignored terms like $-(1/8\pi)\partial (B_\varphi^2)/\partial \varpi$ as being a factor of the disk aspect ratio $\ll 1$ smaller than the
retained term $(1/4\pi)\partial (B_zB_\varpi)/\partial z$ for a spatially thin disk.  Retention of the effects
of magnetic and gas pressure, which generally decrease with increasing radius $\varpi$, would only worsen the
problem of sub-Keplerianity detailed below.

With $\bar f$ denoting the $z$-averaged fraction of
Keplerian rotation at $\varpi$, $\Omega$ $\equiv$ $\bar f\Omega_{\rm K}$ where $\Omega_{\rm K}$ $\equiv$ 
$(GM_*/\varpi^3)^{1/2}$, we obtain (S07, Appendix C)
\be
1-\bar f^2 = {\varpi^2 B_zB_\varpi^+\over 2\pi GM_*\Sigma} .
\label{departure}
\ee
Magnetic effects also affect the vertical structure of the disk.  We define a characteristic thermal velocity
$a$ such that $a^2 \equiv 2z_0P(0)/\Sigma$, where $P(0)$ is the midplane gas pressure.
Integrating the equation of vertical hydrostatic equilibrium in $z$
gives (S07, Appendix C):
\be
{(B_\varpi^+)^2\over 2\pi} = \Sigma\left({2a^2\over A\varpi}-{GM_*A\over \varpi^2}\right),
\label{verthyd}
\ee
where we define $A$ and $A_0$ as, respectively, the actual and thermal aspect ratios of the disk, 
\be
A \equiv {z_0\over \varpi}, \qquad A_0 \equiv \left( {2a^2\varpi\over GM_*}\right)^{1/2}.
\label{A0}
\ee
In the above, $z_0$ is a vertical height of the disk defined so that
the $z$-integrated value of $z\rho$ from the midplane to the surface of zero gas-pressure equals $(z_0/2)(\Sigma/2)$.   Note that $A_0 = \sqrt{2}(a/\varpi\Omega_K)\ll 1$ in a thin disk where $\varpi\Omega_K$ is the Kepleran speed at radius $\varpi$.

WK93, FP95, F97, CF00, and SKW07 never employ the relation (\ref{departure}) to estimate $\bar f$.  
Instead, they allow departures from hydrostatic equilibrium above the disk that involve
advective accelerations in the radial and vertical directions that are ${\cal O}(A_0)$ when measured in a non-dimensional sense.  Section 4 demonstrates that such departures occur only if there is a transonic diffusion of gas across magnetic field lines
that include a strong component of $B_\varphi$ not present in the analysis of S07.

Denoting $B_\varpi^+/B_z = I_\ell$ with $B_z \propto \varpi^{-(1+\ell)}$ in self-similar models,
and substituting equation (\ref{verthyd}) into equation (\ref{departure}), we get
\be
1-\bar f^2 = {A_0\over I_\ell }\left( {A_0\over A}-{A\over A_0}\right).
\label{net}
\ee
The presence of a radial field $B_\varpi$ that increases from zero in the midplane to $B_\varpi^+$ at the surface compresses the disk and makes $A$ smaller than $A_0$.  In the process, $1-\bar f^2$ becomes greater than zero.  

To make contact with disk-wind theory, let us follow FP95 in defining a ``magnetization,''
\be
\mu \equiv {B_z^2\over 4\pi P(0)}= {B_z^2 z_0\over 2\pi \Sigma a^2},
\label{mu}
\ee
which also equals a quantity that WK93 call $\alpha_0^2$.
Dividing equation (\ref{verthyd}) by equation (\ref{mu}) now yields
$I_\ell^2$ = $(2/\mu)( 1-A^2/A_0^2)$,
where we have replaced $B_\varpi^+/B_z$ by $I_\ell$. 
Using this relation to eliminate $A/A_0$ from equation (\ref{net}) now gives
\be
1-\bar f^2 = {A_0\over I_\ell}\left[{1\over C_\ell (\mu)}-C_\ell(\mu)\right],
\label{avdeparture0}
\ee
where the positive quantity $C_\ell (\mu)=A/A_0$ is $C_\ell(\mu)$ $\equiv$ $(1 -I_\ell^2\mu/2)^{1/2}$ $\le$ 1.

The maximum value of $\mu$ allowed mechanically
in equation (\ref{avdeparture0}) comes by setting $\bar f = 0$.   
The aspect ratio then assumes the minimum value:
\be
A_{\rm mech} = {I_\ell\over 2}\left[ \sqrt{1+4\left({A_0/ I_\ell}\right)^2} -1\right].
\label{Amech}
\ee
From $C_\ell(\mu_{\rm mech}) = A_{\rm mech}/A_0$, we obtain $\mu_{\rm mech}$:
\be
\mu_{\rm mech} = {1\over A_0^2}\left[ \sqrt{1+4\left(A_0/ I_\ell\right)^2} -1\right].
\label{mech}
\ee
For $\mu > \mu_{\rm mech}$, magnetic forces are too strong for centrifugal equilibrium
to be possible. For $\mu = \mu_{\rm mech}$ in thin disks where $A_0/I_\ell \ll 1$, the roots (\ref{Amech}) and (\ref{mech}) take the approximate forms,
$A_{\rm mech} \approx A_0^2/I_\ell$, $\mu_{\rm mech} \approx 2/I_\ell^2$. 

In the model of BP82 and in cases where the wind is
relatively lightly loaded (e.g., F97),
$\ell = 1/4$ or nearly so, i.e., $B_z \propto \varpi^{-5/4}$.  The magnetic configuration just above the
disk is then well approximated by a vacuum field, where the
analysis of S07 yields $I_\ell = 1.43$.  The
corresponding inclination angle of the field from the vertical is $i = 55^\circ$.  
With $I_\ell = 1.43$, $\mu_{\rm mech} \approx 0.978$. 
For $\mu$ not closely approaching $\mu_{\rm mech}$, say, $\mu \approx 0.731$ where the magnetic compression $C(\mu) \approx 0.5$,  the right-hand side of equation (\ref{avdeparture0}) is ${\cal O}(A_0) \ll 1$.   The analogous feature in
numerical calculations, combined with the ameliorating effects of rapid diffusion,
led disk-wind theorists to assert that the departure from Keplerian rotation remains small enough for other effects to overcome the deficit.  

\section{Thermal Launching}

Thermal pressure cannot do the requisite job in the absence of fast (turbulent) diffusion (see \S 4).  Published models have the deficit of $f$ from unity at the compressed surface $z=z_0$ both larger than average (e.g., Fig. 5 of FP95) and smaller than average (e.g., Fig. 1 of SKW07).  As a fiducial case, we assume the disk's surface to have the $z$-averaged $f$ obtained from equation
(\ref{avdeparture0}).  To have a significant pressure exhaust of gas onto open field lines, we then require the thermal speed squared $a_s^2$ at the surface of the disk to be (S07, eq.~31):
\be
a_s^2 = {1\over 4}(1-\bar f^2){GM_*\over \varpi} = 
{1\over 2I_\ell A_0}\left[{1\over C_\ell(\mu)}-C_\ell(\mu)\right]a^2 ,
\label{surfacesound}
\ee
where we applied equations (\ref{A0}) and (\ref{avdeparture0}) in the
last step.  

Without external heating, the surface temperature of the
disk, $\sim 2^{-1/4}$ times the effective temperature, will be lower
than its characteristic interior value, i.e., $a_s^2 =
\Theta_s a^2$ where $\Theta_s$ is a fraction $\lesssim 1$.  Equation
(\ref{surfacesound}) then yields a quadratic equation for $C_\ell(\mu_{\rm therm})=A_{\rm therm}/A_0$, whose solution is
\be
A_{\rm therm} = A_0\left[ \sqrt{1+(I_\ell \Theta_s A_0)^2}- (I_\ell \Theta_s A_0)\right].
\label{Atherm}
\ee
From $C_\ell(\mu_{\rm therm}) = A_{\rm therm}/A_0$, we obtain $\mu_{\rm therm}$:
\be
\mu_{\rm therm} = 4{\Theta_s A_0\over I_\ell}
\left[\sqrt{1+(I_\ell\Theta_s A_0)^2}-(I_\ell\Theta_s A_0)\right].
\label{mutherm}
\ee
  
For thin disks where $I_\ell \Theta_s A_0 \ll 1$, equations (\ref{Atherm}) and (\ref{mutherm}) give
$A_{\rm therm} \approx A_0$, $\mu_{\rm therm} \approx 4\Theta_sA_0/I_\ell$.
The right-hand side of the second relation is, at most, several
percent in the inner tenth of an AU in YSO disks.  The
magnetization is then so weak that the associated departure from
Keplerian rotation from equation (\ref{avdeparture0}),
$1-\bar f^2$ $\approx$ $2\Theta_sA_0^2$,
would be similar to that implied by the radial gradients of the thermal gas pressure, i.e.,
very small.  But weak magnetization, $\mu \le
\mu_{\rm therm}$, with good field loading,
does not provide a long enough lever arm to launch fast disk
winds even if $\bar f=1$ (see \S 3.5 of CF00).

If the disk magnetization $\mu$ is chosen greater than
$\mu_{\rm therm}$, but less than $\mu_{\rm mech}$, the actual departure $1-\bar f^2$ $\sim$ ${\cal O}(A_0)$
exceeds the depth of the dimensionless effective potential $1-\bar f^2$ $\sim$ ${\cal O}(A_0^2)$
that the surface temperature can offset.  Because $a_s^2$ may be less than 1\% of $GM_*/\varpi$ in the inner disk, even $\bar f = 0.99$ may constitute a substantial barrier by these standards.
As argued earlier, no room exists for a compromise $\mu$ between $\mu_{\rm therm}$ and $\mu_{\rm mech}$ that will drive disk winds that both load and fling if one depends on thermal pressure to launch.  

\section{Diffusive Loading}

WK93, FP95, F97, CF00, and SKW07 differ with the analysis of \S\S 2 and 3 because they
do not invoke thermal pressure to launch disk winds.  Instead, they overcome local deficits in
$f$ by diffusive loading onto magnetic-field lines coupled to magnetocentrifugal acceleration.

For an imperfectly conducting, lightly ionized, gas, sufficiently collisional to act as a single-component fluid of velocity $\bf v$, the induction equation reads
\be
{\partial {\bf B}\over \partial t}+\nabla \times \left( {\bf B}\times {\bf v}\right) = -\nabla \times \left[ {\bf \eta}\cdot (\nabla \times {\bf B})\right],
\label{fullinduction}
\ee
where $\bf \eta$ is a generalized tensor resistivity associated with Ohm's law (Norman \& Heyvaerts 1985, see especially eq.~22).   In a frame that corotates locally with the magnetic field, so that $\partial {\bf B}/\partial t = 0,$ we may ``uncurl'' equation (\ref{fullinduction})
and obtain the balance of advection and diffusion, ${\bf B} \times {\bf u}$ = $-{\bf \eta}\cdot (\nabla \times {\bf B})$,
where $\bf u$ is the fluid velocity relative to the rotating field lines.  

Keeping only the dominant terms in a thin disk, we follow WK93 and SKW07 in decomposing the advection-diffusion balance into its $\varphi$ and $\varpi$ components:
\be
B_z u_\varpi -B_\varpi u_z = \eta_2  {\partial B_\varphi \over \partial z}-\eta_1 {\partial B_\varpi \over \partial z},
\label{raddrift}
\ee
\be
B_\varphi u_z-B_z(f- f_B)\sqrt{{GM_*\over \varpi}} = \eta_1 {\partial B_\varphi \over \partial z}-\eta_2 {\partial B_\varpi \over \partial z} ,
\label{tandrift}
\ee
where the dimensionless rotation of the field $f_B$ is a function of $z$ at fixed $\varpi$ (in such an Eulerian description, we are not following a given field line).  For WK93, $\eta_1$ is associated with the coefficient of ambipolar diffusion, and
$\eta_2$, with the inverse of the Hall conductivity.   In F97, and CF00, $\eta_1$ is replaced by $\nu_m$
in equation (\ref{raddrift}) and by $\nu_m^{\,\prime}$ in equation (\ref{tandrift}), with the other, non-diagonal, coefficient $\eta_2$ set to zero.

In the disk proper (where $u_z$ is small) of both Figure 1 of SKW07 and Figure 5 of FP95, the terms on the right-hand sides of equations (\ref{raddrift}) and (\ref{tandrift}) contribute, respectively, negatively to the radial drift and positively to the tangential drift relative to the field lines.  These relative motions reverse as the gas climbs vertically into the disk wind.  Eventually,
the right-hand sides vanish as ideal MHD is approached ($\eta_1$ and $\eta_2$ $\rightarrow 0$), with $\bf u$ becoming
parallel to $\bf B$ in the corotating frame. The physical interpretation of the result is as follows. 

A field line rotates at a dimensionless angular velocity $f_B$ that is intermediate between the rotation rate $f$ of the gas near the midplane layers of the disk and the gas in the upper atmosphere that blows into a disk wind. This rate is fixed in steady state so that angular momentum balance is reached following a given field line (a condition not examined in this paper). In the WK93 description, the neutral gas near the midplane would rotate at the Kepler speed except for collisions with ions and electrons that are tied to the more slowly rotating magnetic field lines.
The collisions therefore make the neutral gas lose angular angular momentum, which then cause the neutrals to drift radially inward relative to the radially stationary field lines and the ions and electrons tied to them.  

Freezing to field lines in the upper atmosphere of the disk yields the familiar effect of a ``sliding bead on a rotating rigid wire'' (Chan \& Henriksen 1980) that accelerates the gas through the slow MHD transition as it gains angular momentum at the expense of the matter in the deeper layers of the disk.  Thus, the gas that participates in the disk wind ends up rotating at super-Keplerian speeds.  A similar phenomenon occurs in the FP95 description, except that the role of collisions between neutrals and ions/electrons is replaced by a mixing of fluid parcels created by MHD turbulence.

When an outwardly bending field line rotates uniformly along its length, there will always be a point
where centrifugal effects on an electrically conducting test particle placed there
counterbalances an opposing gravitational field.  In Figure 5 of FP95 and Figure 1 of SKW07,
this point occurs at $\sim 2v_T/\Omega_K$, where $v_T$ is the thermal velocity in the midplane.  For
a vertically isothermal disk (the most optimistic case), where $a = (2/\sqrt{\pi})v_T$, this point is $\sim 2.5 z_0$ 
since $z_0 \sim 0.5 A_0\varpi \sim 0.8 v_T/\Omega_K$ for typical
parameters used to launch disk winds. The height $z_0$ is defined by where the condition of vertical hydrostatic equilibrium
would have created a surface.  As in any atmosphere, the gas density does not really drop abruptly to
zero at $z_0$, but unless diffusion works very efficiently, there cannot be much gas at $2.5z_0$.

Trouble on this point enters when numerical solutions are adjusted to give large rates of diffusion.  In order of magnitude, let us estimate $\partial B_\varpi/\partial z \sim B_\varpi^+/z_0$ and
$\partial B_\varphi/\partial z \sim B_\varphi^+/z_0$ in the disk proper.  In the more successful cases considered by FP95, the quantities $\eta_1 \sim z_0 a\sim \eta_2$.  With $B_\varpi^+ >0$ and $-B_\varphi^+>0$ comparable in magnitude to $B_z>0$, equations (\ref{raddrift}) and (\ref{tandrift}), with $u_z$ negligible in the disk proper, now yield the estimates,
\be
u_\varpi \sim -a, \qquad f-f_B\sim {a\over \sqrt{GM_*/\varpi}}\sim A_0 .
\label{driftspeeds}
\ee

The first deduction, that the inward drift speed in the bulk of the disk approaches sonic values, is verified, for example, by Figure 4 of FP95 where $u_\varpi \approx -v_T$.    The second deduction, that the dimensionless rotation rate of the gas $f$ can exceed that of the field $f_B$ by ${\cal O}(A_0)$, would produce favorable conditions for the launching of disk winds if the field rotates slower than the gas in the disk proper, but still within fraction ${\cal O}(A_0)$ of Keplerian.

In the work of FP95, the diffusion coefficients $\eta_1$ and $\eta_2$ are chosen large enough so that, unlike S07, the magnetic field is not well coupled to the matter in the deep interior of the disk.  With $\eta_1 \sim z_0 a$, the diffusive time scale to straighten poloidal field lines that are bent on a scale $z_0$ is $\sim z_0^2/\eta_1 \sim z_0/a$, which is of order the inverse Kepler rate $\Omega_{\rm K}^{-1}$ (or shorter, if the disk is severely magnetically compressed).  The dynamical slippage of the matter past the field, both in the $\varpi$ and $\varphi$ directions, is a serious weakness of the models.
   
Rapid diffusion can carry gas to the slow MHD transition point beyond which magnetocentrifugal fling
works in the outward direction to produce an outflowing wind.
Such a strategy generates, through equations (\ref{raddrift}) and (\ref{driftspeeds}), radial drift speeds inside the disk that are transonic $\sim -v_T$.  Consider what this implies if a disk wind were responsbile for the primary accretion mechanism of the disk all the way out to an outer disk radius $\sim 300$ AU in a low-mass YSO.  The midplane temperature at 300 AU is unlikely to be much less than, say, 20 K, which corresponds to an isothermal sound speed in cosmic molecular gas of $v_T = 0.278$ km s$^{-1}$. The
whole disk would then shrink onto the star in 300 AU/(0.278 km s$^{-1}) \sim$ 5,000 yr, which is far too short a time scale
to be realistic.  

If global magnetorotational instability is a viable alternative for YSO accretion disks, the analysis of S07 and Shu et al.~(2008) implies that the turbulent resistivity $\eta_1$ is smaller than estimated by FP95 or F97 by an additional factor $\sim \mu A \ll 1$.  If we continue to assume that $-B_\varphi^+$ and $B_\varpi^+$ are comparable to $B_z$, with the former holding only if disk winds can be launched, then equations (\ref{raddrift}) and (\ref{tandrift}) imply
\be
u_\varpi \sim -\mu A a, \qquad f-f_B \sim \mu AA_0\lesssim A_0^2.
\ee
The second relation above states that diffusion can now make up deficits of $f_B$ only by amounts $\lesssim A_0^2$, where $f_B$ itself is given by $\bar f$ to ${\cal O}(A_0^2)$ in such circumstances.  In turn, $\bar f$ departs from Keplerian rotation typically by ${\cal O}(A_0)$ for strongly magnetized disks according to equation (\ref{avdeparture0}).
A low (S07; see also Lubow et al.~1993) rather than a high (FP95 or F97) rate of field diffusion yields equation (\ref{surfacesound}) as, after all, the correct approximate criterion for wind loading and launching. 

The numerical choices made by WK93 are more conservative than those of FP95, and the net results for $u_\varpi$ and $f-f_B$ are intermediate between the high- and low-diffusion scenarios described above (see Appendix of WK93). Specifically, in the numerical case depicted in Figure 1 of SKW07, where $\alpha_0 \equiv \sqrt{\mu} = 0.95$ and $a_s^2 = v_T^2 = 0.01 GM_*/\varpi$, the dimensionless loading is $\kappa = 3.2\times 10^{-4}$ and the square of the lever arm is $\lambda = 395$ (these are $\beta^{-1}$ and $J_w$, respectively, in the notation of X-wind theory), corresponding indeed to a disk wind that can fling, but is very lightly loaded.  Such winds do not describe observed YSO jets well.  The order of magnitude changes (exponential sensitivity) in quantities like $\kappa$ and $\lambda$ when $\eta$ changes by a factor of a few gives powerful testimony to the crucial role that magnetic diffusivity plays in current theories of the launching of disk winds. 

\section{Discussion and Conclusion}


Disk winds have been invoked as a solution to the angular momentum problem
in accretion disks (e.g., Pelletier \& Pudritz 1992).  The challenge of sub-Keplerian
rotation facing magnetocentrifugally-driven, cold, disk winds for much, if not most, of
its radial span suggests that internal mechanisms other than the
back reaction to disk winds must account for the primary accretion
mechanism of such disks.  Examples of such mechanisms include
spiral density waves for high-mass disks and the magnetorotational instability
for low-mass disks. 

Disruption of the disk at an inner edge by a stellar magnetosphere
will drive an X-wind (Cai et al.~2008 and references therein) if the resistivity in the disk is appreciably smaller than the viscosity
(for numerical simulations, see Romanova et al. 2008).
How does X-wind theory escape the
conundrum posed by equation (\ref{centrifugal})?  Near the inner edges
of disks, the X field geometry involves a swing of outwardly bending field lines (right half of X)
to inwardly bending field lines (left half of X) as accretion disks are truncated by
funnel flows through their interactions with a stellar magnetosphere.
The swing of $B_\varpi^+B_z$ from positive values through
zero (vertical field lines that are ``dead'' to outflow or inflow)
automatically promotes a transition from sub-Keplerian to
Keplerian rotation.  Indeed, when the density decreases sharply inward on a fractional radial scale $\sim A_0$, the no-longer ignorable, extra, radial push of
gas pressure toward the inner edge speeds up disk rotation and makes the footpoints of some of the outwardly bending
field lines exterior to the dead zone sufficiently close to Keplerian rotation
as to enable the thermal launching of X-winds (see the discussion of Shu et al.~2008). 

\acknowledgments
We thank Sylvie Cabrit, Jonathan Ferreira, Al Glassgold, Arieh K\"onigl, Raquel Salmeron,  and Mark Wardle for comments that improved the presentation of the paper.
FS acknowledges support from the Physics Department of UCSD; SL from
CONACyT 48901 and PAPIIT-UNAM IN106107; DG, from INAF-OAA and EC grant MRTN-CT-2006-035890;
MC, from a grant from the NSC to TIARA; SM, from a Spitzer Fellowship.


\begin{thebibliography}{}
\bibitem[Blandford \& Payne(1982)]{Blandford:1982}
Blandford, R. D., Payne, D. G. 1982 MNRAS, 199, 883 (BP82)
\bibitem[]{}
Cai, M., Lin, H. H., Shang, H., Shu, F. H. 2008, ApJ, 672, 489
\bibitem[]{}
Chan, K. L.,  Henriksen, R. N. 1980, ApJ. 241, 534
\bibitem[]{}
Casse, F., Ferreira, J. 2000, A\&A, 353, 1115 (CF00)
\bibitem[]{}
Ferreira, J. 1997, A\&A, 319, 340 (F97)
\bibitem[]{}
Ferreira, J., Pelletier, G. 1995, 295, 807 (FP95)
\bibitem[]{}
Lubow, S. H., Papaloizou, J., Pringle, J. E. 1994, MNRAS, 267, 1214
\bibitem[]{}
Norman, C., Heyvaerts, J. 1985, A\&A, 147, 247
\bibitem[]{}
Pelletier, G., Pudritz, R. E. 1992, ApJ, 394, 117
\bibitem[]{}
Pudritz, R. E., Norman, C. A., ApJ, 274, 677
\bibitem[]{}
Romanova, M. M., Long, M., Kulkarni, A. K., Kurosawa, R., Ustyugova, G. V., Koldoba, A. K., Lovelace, R. V. E. 2008, in IAU Symp. Proc. 243, Star-Disk Interaction in Young Stars, ed. I. Appenzeller \& J. Bouvier (Kluwer), in press
\bibitem[]{}
Salmeron, R., K\"onigl, A., Wardle, M.~2007, MNRAS, 375, 177
\bibitem[]{}
Shu, F. H., Galli, D., Lizano, S., Cai, M. 2008, in IAU Symp. Proc. 243, 
Star-Disk Interaction in Young Stars, ed. I. Appenzeller \& J. Bouvier (Kluwer), in press
\bibitem[]{}
Shu, F. H., Galli, D., Lizano, S., Glassgold, A. E., Diamond, P. 2007, ApJ, 665, 
535 (S07)
\bibitem[]{}
Uchida, Y., Shibata, K. 1985, PASJ, 37, 31
\bibitem[]{}
Wardle, M., K\"onigl, A. 1993, ApJ, 410. 218 (WK93)

\end{thebibliography}
\end{document}